\documentclass[twocolumn]{aastex62}

\received{\today}

\shortauthors{Bandyopadhyay et al.}

\begin{document}
	
\title{Incompressive Energy Transfer in the Earth's Magnetosheath: Magnetospheric Multiscale Observations}

\author[0000-0002-6962-0959]{Riddhi Bandyopadhyay}
\email{riddhib@udel.edu}
\affiliation{Bartol Research Institute and Department of Physics and Astronomy, University of Delaware, Newark, DE 19716, USA}
	
\author[0000-0001-8478-5797]{A. Chasapis}
\affiliation{Bartol Research Institute and Department of Physics and Astronomy, University of Delaware, Newark, DE 19716, USA}
	
\author[0000-0002-7174-6948]{R. Chhiber}
\affiliation{Bartol Research Institute and Department of Physics and Astronomy, University of Delaware, Newark, DE 19716, USA}
	
\author[0000-0003-0602-8381]{T.~N. Parashar}
\affiliation{Bartol Research Institute and Department of Physics and Astronomy, University of Delaware, Newark, DE 19716, USA}
	
\author[0000-0001-7224-6024]{W.~H. Matthaeus}
\email{whm@udel.edu}
\affiliation{Bartol Research Institute and Department of Physics and Astronomy, University of Delaware, Newark, DE 19716, USA}
	
\author[0000-0003-1861-4767]{M.~A. Shay}
\affiliation{Bartol Research Institute and Department of Physics and Astronomy, University of Delaware, Newark, DE 19716, USA}
	
\author[0000-0002-2229-5618]{B.~A. Maruca}
\affiliation{Bartol Research Institute and Department of Physics and Astronomy, University of Delaware, Newark, DE 19716, USA}

\author[0000-0003-0452-8403]{J.~L. Burch}
\affiliation{Southwest Research Institute, San Antonio, TX, USA}

\author[0000-0002-3150-1137]{T.~E. Moore} 
\affiliation{NASA Goddard Space Flight Center, Greenbelt, MD, USA}

\author[0000-0001-9249-3540]{C.~J. Pollock}
\affiliation{Denali Scientific, Fairbanks, Alaska, USA}

\author[0000-0001-8054-825X]{B.~L. Giles}
\affiliation{NASA Goddard Space Flight Center, Greenbelt, MD, USA}

\author{W.~R. Paterson}
\affiliation{NASA Goddard Space Flight Center, Greenbelt, MD, USA}

\author{J. Dorelli}
\affiliation{NASA Goddard Space Flight Center, Greenbelt, MD, USA}

\author[0000-0003-1304-4769]{D.~J. Gershman}
\affiliation{NASA Goddard Space Flight Center, Greenbelt, MD, USA}

\author{R.~B. Torbert}
\affiliation{University of New Hampshire, Durham, NH, USA}

\author[0000-0003-1639-8298]{C.~T. Russell}
\affiliation{University of California, Los Angeles, CA, USA}

\author[0000-0001-9839-1828]{R.~J. Strangeway}
\affiliation{University of California, Los Angeles, CA, USA}

	
	
\begin{abstract}
Using observational data from the \emph{Magnetospheric Multiscale} (MMS) Mission in the Earth's magnetosheath, we estimate the energy cascade rate at three 
ranges of length scale, 
using different techniques within the framework of incompressible magnetohydrodynamic (MHD) turbulence. At the energy containing scale, the energy budget is controlled by the von K\'arm\'an decay law. Inertial range cascade is estimated by fitting a linear scaling to the mixed third-order structure function. Finally, we use a multi-spacecraft technique to estimate the  Kolmogorov-Yaglom-like cascade rate in the kinetic range, well below the ion inertial length scale, where we expect a reduction due to involvement of other channels of transfer. The computed 
inertial range cascade rate is almost equal to the von K\'arm\'an-MHD law at the energy containing scale, while the incompressive cascade rate evaluated at the kinetic scale is somewhat lower, as anticipated in 
theory~\citep{Yang2017PoP}. In agreement with a recent study~\citep{Hadid2018PRL}, we find that the incompressive cascade rate in the Earth's magnetosheath is about $1000$ times larger than the cascade rate in the pristine solar wind. 
\end{abstract}



\section{Introduction} \label{sec:intro}

One of the long standing mysteries of space physics is the anomalous heating of the solar wind. Assuming adiabatic expansion, the temperature profile of the solar wind is expected to scale as $T(r) \sim r^{-4/3}$, where $r$ is the radial distance from the Sun. Yet, the best fit to Voyager temperature observation~\citep{Richardson1995GRL}
results in a radial profile $T(r) \sim r^{-1/2}$. Turbulence provides a natural explanation, supplying internal energy through a 
cascade process that channels
available energy, in the form of electromagnetic fluctuations and velocity shear at large scales, to smaller scales and ultimately into 
dissipation and heating. In collisionless plasmas, such as the solar wind or the magnetosheath, the situation is more complicated due to kinetic effects. Nevertheless, magnetohydrodynamics, which models the plasma as a single fluid, has proven to be a very successful theoretical framework in describing even weakly collisional
plasmas, such as the solar wind, provided one focuses on 
large-scale features and processes. 
In the last few decades, there have been extensive studies related to energy cascade rate and dissipation channels in collisionless plasmas 
~\citep{MacBride2008ApJ,Sorriso-Valvo2007PRL,Osman2011PRL,Coburn2014PRS},
largely based on ideas originating in MHD studies ~\citep{Politano1998GRL}. 
Recently, there has been 
an effort to understand the more complex 
pathways of energy cascade in 
plasmas~\citep{Howes2008JGR,Yang2017PoP,Yang2017PRE,Servidio2017PRL,Hellinger2018ApJL}. 
Prior to presenting new observational results on this timely subject, 
to provide context, we now briefly digress on this history.  

\citet{Taylor1935} suggested, based on heuristic arguments, 
that the decay rate in a neutral fluid is controlled by the energy containing eddies. 
Later,~\citet{Karman1938PRSL} derived Taylor's results more rigorously, assuming that the shape of the two-point correlation function remains unchanged during the decay of a turbulent fluid at high Reynolds number. In one of his three famous 1941 papers,~\citet{Kolmogrov1941c} derived 
an exact expression for the inertial range cascade rate -- 
the so-called third-order law for homogeneous, incompressible, isotropic neutral fluids 
-- from the Navier-Stokes equation, based on only few general assumptions.  ~\citet{Hossain1995PoP} attempted to extend von K\'arm\'an phenomenology for magnetized fluids based on dimensional arguments. Politano and Pouquet~\citep{Politano1998GRL,Politano1998PRE}(PP98) extended Kolmogorov's \emph{third-order} law to homogeneous, incompressible MHD turbulence using Elsasser variables. 

Following these theoretical advances, 
the third-order law has been used in several studies to estimate the energy cascade rate in the solar wind~\citep{Sorriso-Valvo2007PRL,MacBride2008ApJ,Marino2012ApJ,Coburn2012ApJ,Coburn2014ApJ,Coburn2014PRS}. Density fluctuations in the solar wind are usually low enough so that incompressible MHD works well. \cite{Carbone2009PRL} and~\cite{Carbone2012SSR} first made an attempt, based on  heuristic reasoning, to include density fluctuations for estimating compressible transfer rate using the ``third-order" law. Recently, ~\citet{Banerjee2013PRE} (BG13) worked out an exact transfer rate for a compressible medium. Hadid \textit{et al.}(~\citeyear{Hadid2015ApJL,Hadid2018PRL}) compared the cascade rates derived from BG13 and PP98 in weakly and highly compressive media in various planetary magnetosheaths and solar wind~\citep{Banerjee2016ApJL}. It was found that the fluxes derived from the two theories lie close to each other for weakly compressive media and the deviation starts to become significant as the plasma compressibility becomes higher, as expected. 

Parallel to observational works in space plasma systems, on the theoretical side there have been numerous efforts to refine the ``third-order'' law derived for incompressible homogeneous MHD, by including more kinetic physics, like two-fluid MHD~\citep{Andres2016PRE}, Hall MHD~\citep{Andres2018PRE}, electron MHD~\citep{Galtier2008JGR}, by  incorporating the effect of large-scale shear, slowly varying mean field~\citep{Wan2009PoP,Wan2010aPoP}, etc. One would expect, as kinetic effects become important in a plasma, such corrections would need to be taken into account. In this work, we consider only incompressible, homogeneous MHD phenomenologies.

In this study, we focus on the Earth's magnetosheath. While similar to the turbulence observed in the pristine solar wind, the shocked solar wind plasma in the magnetosheath, downstream of Earth's bow-shock, provides a unique laboratory for the study of turbulent dissipation under a wide range of conditions, like plasma beta, particle velocities, compressibility etc. Past studies, both numerical~\citep{Karimabadi2014PoP} and observational~\citep{Sundkvist2007PRL,Huang2014ApJL,Chasapis2015ApJL,BreuillardApJ2016,Yordanova2016GRL,Chasapis2017ApJ} have probed the properties of turbulent dissipation in the magnetosheath. Such studies have established the contribution of intermittent structures, such as current sheets to turbulent dissipation in the kinetic range. However, the properties of turbulence at kinetic scales and quantitative treatments of energy dissipation at those scales remain scarce~\citep{Huang2017ApJL,Hadid2018PRL,Gershman2018PoP,Breuillard2018ApJ}. 

Here, we investigate the energy transfer at several scales, including 
kinetic scales, using the state-of-the-art 
capabilities of the the Magnetospheric Multiscale (MMS) 
Mission~\citep{Burch2016SSR}. The combination of high-time-resolution plasma data with multi-spacecraft observations at very small separations allows us to carry out estimation
of the cascade rate using several strategies employing a single data interval.  
For steady conditions, we 
expect that the decay rate of the energy-containing eddies at large scales, $\epsilon_1$ 
and the cascade rate measured in the inertial range, $\epsilon_2$, will be comparable. 
In the subproton kinetic range, additional channels of transfer and dissipation 
become available (e.g., \citet{Yang2017PoP,Yang2017PRE})
and the measured incompressive
cascade rate $\epsilon_3$ at such small scales 
is likely to represent a smaller fraction of the 
total steady transfer rate $\epsilon$. Thus, we expect that
$\epsilon_1 \approx \epsilon_2 > \epsilon_3$ for the measured rates.  
Throughout 
the paper, we indicate the total (expected) cascade rate as $\epsilon$
and the measured values at different scales with a 
suffix.

\section{Data Selection and Overview} \label{sec:overview}

\begin{figure*}
	\begin{center}
		\includegraphics[scale=0.5]{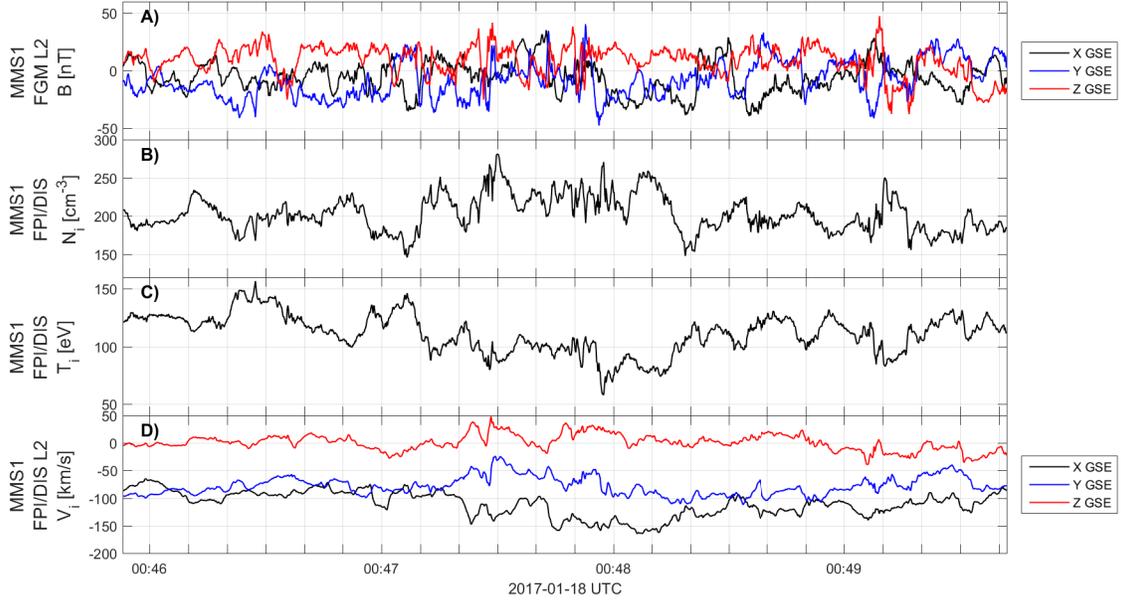}
		\caption{
Overview of the MMS observations in magnetosheath turbulence selected for this study. The data shown is from the FGM and FPI instruments on-board the MMS1 spacecraft. Panel A) shows the magnetic field measurements in GSE coordinates. Panel B) shows the ion density. Panel C) shows the ion temperature. Panel D) shows the ion velocity in GSE coordinates.}
		\label{fig:overview}
	\end{center}
\end{figure*}

\begin{figure}
	\begin{center}
		\includegraphics[scale=1.0]{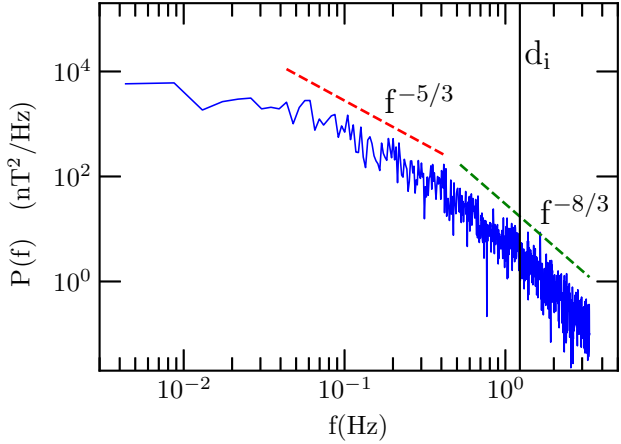}
		\caption{Spectral power density of magnetic field measured by MMS1. Kolmogorov scaling $\sim f^{-5/3}$ is shown for reference. The solid vertical line represents $k d_i = 1$ with the wave vector $k \simeq (2\pi f)/|\langle \mathbf{V} \rangle|$. }
		\label{fig:spec_b}
	\end{center}
\end{figure}

We use burst resolution MMS data obtained in the turbulent magnetosheath on 2017 January 18 from 00:45:53 to 00:49:43 UT. An overview of the interval is shown in Fig.~\ref{fig:overview}.
During this time period, a clear Kolmogorov scaling $(\sim f^{-5/3})$ can be seen in the magnetic energy spectra (See Fig.~\ref{fig:spec_b}). A break in spectral slope from $\sim f^{-5/3}$ to $\sim f^{-8/3}$ is observed near 0.5Hz. Some important plasma parameters of the selected turbulent interval are reported in Table~\ref{tab:overview}. The density fluctuations and the turbulent Mach number are low (see Table~\ref{tab:overview}), similar to those commonly observed in the pristine solar wind, justifying the applicability of an incompressible MHD approach for the selected interval.
Contrary to the pristine solar wind, the ratio of root mean squared (rms) fluctuations to the mean is high for the magnetic field, indicating isotropy, thus making this interval suitable for this study since we only consider incompressible, isotropic MHD here. 

The separation of the four MMS spacecraft was $\approx 8$ km, which corresponds to about half the ion inertial length ($d_i \approx 16$ km).
We used MMS burst resolution data which provides magnetic field measurements (FGM) at $128$ Hz~\citep{Torbert2016SSR,Russell2016SSR}, and ion density, temperature and velocity (FPI) at $33$Hz~\citep{Pollock2016SSR}.
The small separation, combined with the high time resolution of the measurements of the ion  moments allow us to use a multi-spacecraft approach similar to the one used by~\citet{Osman2011PRL} at the very small scales of the turbulent dissipation range. 


\begin{table*}[ht!]
	\caption{Description of some plasma parameters} 
	\label{tab:overview}
	\begin{center}
	\begin{tabular}{c c c c c c c  c c c c c}
		\hline \hline
		$|\langle \mathbf{B} \rangle|$ & $\delta B /|\langle \mathbf{B} \rangle|$ 
		& $\langle n_e \rangle$ & $\delta n_e /\langle n_e \rangle$ 
		& $\langle n_i \rangle$ & $\delta n_i /\langle n_i \rangle$
		& $d_{i}$ & $d_{e}$ 
		&$|\langle \mathbf{V} \rangle|$ & $\delta V /|\langle \mathbf{V} \rangle|$ 
		 & $M_t$ & $\beta_p$ \\
		 \colhead{($\rm nT$)} & \colhead{} & \colhead{(${\rm cm^{-3}}$)} & \colhead{} & \colhead{(${\rm cm^{-3}}$)} & \colhead{} & \colhead{($\rm km$)} & \colhead{($\rm km$)} & \colhead{($\rm km~s^{-1}$)} & \colhead{} & \colhead{} & \colhead{} \\
		\hline
		 13.1 & 1.9 & 169 & 0.11 & 202 & 0.12 & 17.5 & 0.4 & 135 & 0.42 & 0.2 &13\\
		\hline
	\end{tabular}
		\end{center}
	\tablecomments{Data obtained from MMS1 on 2017 January 18 from 00:45:53 to 00:49:43 UT. rms fluctuation amplitude is defined as $\delta B = \sqrt{\langle |\mathbf{B}(t) - \langle \mathbf{B} \rangle|^2 \rangle}$ and similarly for other quantities. Ion inertial length, $d_i $, electron inertial length, $d_e$, ion velocity, $\mathbf{V}$, turbulent Mach number, $M_t = \delta V/v_{th}$, and the proton plasma beta, $\beta_p = v_{th}^2/V_{A}^2$ are also reported.}
\end{table*}

\section{Energy containing scale: $\epsilon_{1}$} \label{sec:eps1}

It is reasonable to expect that the
global decay is controlled, to a suitable level of approximation, 
by von K\'arm\'an decay law, generalized to MHD~\citep{Hossain1995PoP,Politano1998GRL,Politano1998PRE,Wan2012JFM}, 
\begin{eqnarray}
 \epsilon_1 = -\frac{d (Z^{\pm})^2}{d t} = \alpha_{\pm} \frac{(Z^{\pm})^2 Z^{\mp}}{L_{\pm}}\label{eq:eps1},
\end{eqnarray}
where $\alpha_{\pm}$ are positive constants and $Z^{\pm}$ are the rms
fluctuation values of the 
Els\"asser variables defined as
\begin{eqnarray}
\mathbf{Z}^{\pm}(t) = \mathbf{V}(t) \pm \frac{\mathbf{B}(t)}{\sqrt{\mu_{0} m_{p} n_{i}(t)}}\label{eq:Z},
\end{eqnarray}
where the local mean values have been subtracted
from ${\bf V}$ and $\bf B$, respectively
the plasma (ion) velocity and the magnetic field vector.
Here, $\mu_0$ is the magnetic permeability of vacuum, $m_{p} \gg m_{e}$ 
are proton and electron mass, respectively and $n_{i}$ is the number 
density of protons. 

The similarity length scales $L_{\pm}$ 
appearing in Eq.\ref{eq:eps1} are related to the characteristic scales of the 
``energy containing" eddies. Usually, a natural choice for the similarity scales are 
the associated correlation lengths, 
computed from the two-point correlation functions.

The procedure for estimating the required correlation 
lengths is not unique, and in real data environments numerous 
issues may arise \citep{MatthaeusEA99}.
The basis of the estimate is determination of the 
two-point, single-time correlation tensor, 
which 
under suitable conditions -
the Taylor ``frozen-in'' flow hypothesis~\citep{Taylor1938PRSLA,Jokipii1973ARAA} -
is related to the two-time correlation at the spacecraft position. 
The trace of the correlation tensors, computed from 
the Elsasser variables, is defined as
\begin{eqnarray}
R^{\pm} (\tau) = \langle \mathbf{Z^{\pm}}(t) \cdot \mathbf{Z^{\pm}}(t+\tau) \rangle_{T} \label{eq:Rzpm}.
\end{eqnarray}
Here, $\langle \cdots \rangle_{T}$ 
is a time average, usually over 
the total time span of the data. 
We have used the standard Blackman-Tukey method, 
with subtraction of the local mean, to evaluate equation~(\ref{eq:Rzpm}). 
Although the standard definition of correlation scale 
is given by an integral over the correlation function, 
in practice, especially when there is substantial 
low frequency power present, it is advantageous to 
employ an alternative ``1/e'' definition \citep{SmithEA01}, 
namely 
\begin{eqnarray}
R^{\pm}(\tau^{\pm}) = \frac{1}{e}\label{eq:corrtau},\\
L_{\pm} = |\langle \mathbf{V}\rangle| \tau^{\pm}\label{eq:corrL},
\end{eqnarray}
where the second line exploits the Taylor hypothesis. 
Qualitatively, for some well-behaved spectra,
the reciprocal correlation length corresponds to the low frequency 
``break" in the inertial range power law. 
However, estimation of correlation scale by 
identification with the 
break in the spectrum may not be very accurate. 
Equation~(\ref{eq:corrL}) is expected to be 
a more quantitative approximation. 
Furthermore, determination of
the correlation scale is inherently
difficult due to low frequency power.
For example, 
correlation lengths systematically 
increase as the length of the data interval is increased \citep{IsaacsEA15}. 
Use of the $1/e$ method, as seen in equation (\ref{eq:corrtau}), 
(\ref{eq:corrL}) mitigates this sensitivity\citep{SmithEA01}. 

\begin{figure}
	\begin{center}
		\includegraphics[scale=1.0]{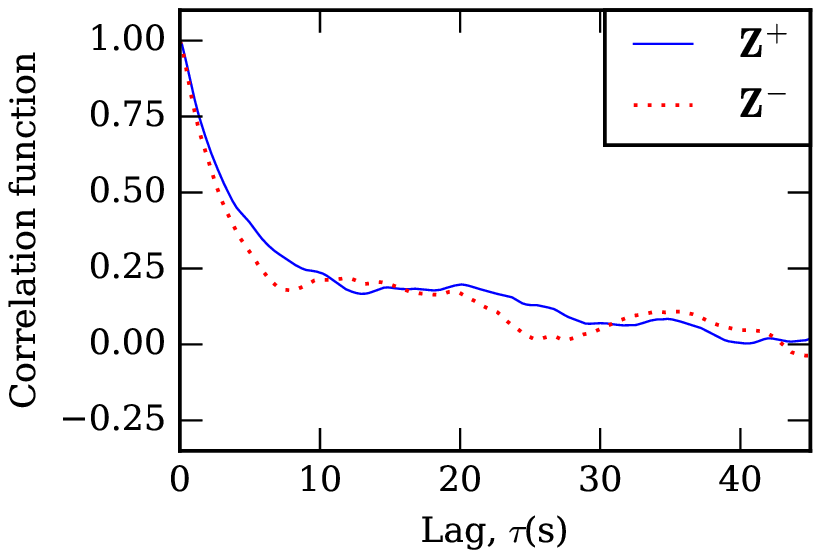}
		\caption{Normalized correlation function vs. time lag (seconds) 
for the Elsasser variables, derived from the measurements obtained from MMS1.
Spatial lags $L$ may be estimated using the Taylor hypothesis with solar wind speed (Table~\ref{tab:overview}) $135$ km/s [See equation~(\ref{eq:corrL})].}
		\label{fig:corr}
	\end{center}
\end{figure}

With these conventions, we first calculate $\mathbf{Z}^{\pm}$ based on ion velocity. We calculate normalized correlation functions for maximum lag of 1/5\textsuperscript{th} of the total dataset. We show the plots of normalized correlation function for each Elsasser variable in Fig.~\ref{fig:corr}. Fitting an exponential function to each of the normalized correlation function gives correlation time $\tau^{+} = 5.6~\mathrm{s}$ and $\tau^{-} = 4~\mathrm{s}$. Note that these magnetosheath correlation times are much shorter than the analogous time scales in the ambient solar wind. While the magnetosheath plasma originates in the solar wind, during transmission through the shock region, it is evidently modified significantly, to accommodate a much smaller ``system size", and the mechanism of driving.

\begin{table}
	\caption{Derived variables}
	\label{tab:zpm}
	\begin{center}
	\begin{tabular}{c c c c c c c}
		\hline \hline
		S/C 
		& $ Z^{+} $ & $L_{+}$ 
		& $Z^{-}$ & $L_{-}$ & $\sigma_c$ 
		 \\
		 & \colhead{($\rm km~s^{-1}$)} &  \colhead{($\rm km$)} & \colhead{($\rm km~s^{-1}$)}  & \colhead{($\rm km$)}   
		\\
		\hline
		MMS1 & 55 & 756 & 42 & 547 & 0.24 \\
		MMS2 & 53 & 608 & 43 & 526 & 0.21 \\
		MMS3 & 53 & 608 & 43 & 526 & 0.21 \\
		MMS4 & 53 & 587 & 44 & 526 & 0.21 \\
		\hline
	\end{tabular}
	\end{center}
	\tablecomments{Els\"asser amplitudes $Z^{\pm}$, correlation lengths $L_{\pm}$,
and normalized cross helicity defined as $\sigma_c = [(Z^{+})^2-(Z^{-})^2]/[(Z^{+})^2+(Z^{-})^2]$.}
\end{table}

In Table~\ref{tab:zpm}, we report the required statistics 
obtained for the Els\"asser variables for this interval.
Putting these values in equation~(\ref{eq:eps1}), we find
\begin{eqnarray}
\epsilon^{+}_1 &=& \alpha_{+} \frac{(Z^{+})^2 Z^{-}}{L_{+}} \nonumber\\
				 &=& \alpha_{+} 168 \times 10^{6} \mathrm{~J~kg^{-1}~s^{-1}} \label{eq:eps1p}.\\
\epsilon^{-}_1 &=& \alpha_{-} \frac{(Z^{-})^2 Z^{+}}{L_{-}} \nonumber\\
&=& \alpha_{-} 177 \times 10^{6} \mathrm{~J~kg^{-1}~s^{-1}} \label{eq:eps1m}.
\end{eqnarray}

The values of the von K\'arm\'an constants $\alpha_{\pm}$ are required to proceed further. The values of the constants are expected to be of order unity. Following Appendix B of ~\citet{Usmanov2014ApJ},  for isotropic and low cross helicity case, $\alpha = 4 C_{\epsilon}/ (9\sqrt{3})$, where $C_{\epsilon}$ is the dimensionless dissipation rate. In ~\cite{Usmanov2014ApJ}, it was assumed $C_{\epsilon} \simeq 0.5$ because that is the value found for fluid turbulence~\citep{Pearson2002PoF, Pearson2004PoF}. Recent investigations show that in MHD the value of $C_{\epsilon}$ is quite low compared to the fluid value. In a series of papers Linkmann et al.~(\citeyear{Linkmann2015PRL, Linkmann2017PRE}) showed that for isotropic, low cross helicity MHD, $C_{\epsilon} \simeq 0.265$. Therefore, for low cross helicity, $C_{\epsilon}^{+} \simeq C_{\epsilon}^{-} \simeq 0.133$. Using this value, we obtain 
$\alpha_{+} \simeq \alpha_{-} \simeq 0.03$. Inserting 
these in equations~(\ref{eq:eps1p}) and~(\ref{eq:eps1m}) for MMS1 data in Table~\ref{tab:zpm} we get
\begin{eqnarray}
\epsilon^{+}_1 &\simeq& 5.0 \times 10^{6} \mathrm{~J~kg^{-1}~s^{-1}} \label{eq:eps1p_val}.\\
\epsilon^{-}_1 &\simeq& 5.3 \times 10^{6} \mathrm{~J~kg^{-1}~s^{-1}} \label{eq:eps1m_val}.
\end{eqnarray}
We perform the same calculation for all four spacecraft listed in Table~\ref{tab:zpm}. Obtained values of $\epsilon^{\pm}_1$ are reported in Table~\ref{tab:eps1}.
\begin{table}
	\caption{Global decay rate estimates from von K\'arm\'an law}
	\label{tab:eps1}
	\begin{center}
	\begin{tabular}{c c c}
		\hline \hline
		S/C 
		& $ \epsilon^{+}_{1} $ & $ \epsilon^{-}_{1} $  
		\\
		& \colhead{($\rm J~kg^{-1}~s^{-1}$)} &  \colhead{($\rm J~kg^{-1}~s^{-1}$)}    
		\\
		\hline
		MMS1 & $5.0\times10^6$ & $5.3\times10^6$  \\
		MMS2 & $6.0\times10^6$ & $5.6\times10^6$  \\
		MMS3 & $6.0\times10^6$ & $5.6\times10^6$  \\
		MMS4 & $6.3\times10^6$ & $5.9\times10^6$  \\
		Average & $(5.8\pm0.5)\times10^6$ & $(5.6\pm0.2)\times10^6$ \\
		\hline
	\end{tabular}
	\end{center}
	\tablecomments{The last row is the average of all four spacecraft measurements listed in the first four rows. The uncertainty is the  standard deviation of the four measurements.}
\end{table}

\section{Inertial range: $\epsilon_{2}$} \label{sec:eps2}
To estimate $\epsilon_2$, 
the energy transfer rate in the 
inertial scale, we use Kolmorogov-Yaglom law, extended to isotropic MHD,
\begin{eqnarray}
Y^{\pm}(r) = - \frac{4}{3} \epsilon^{\pm} r \label{eq:3ord},
\end{eqnarray}
where $Y^{\pm}(r)=\langle \mathbf{\hat{r}}\cdot \Delta \mathbf{Z}^{\mp}(\mathbf{r}) |\Delta \mathbf{Z}^{\pm}(\mathbf{r})|^2 \rangle$, are the mixed third-order structure functions. Equation~(\ref{eq:3ord}) has been the standard approach in estimating the inertial range energy transfer rate in the solar wind although it is clearly not an isotropic system $(\delta B /|\langle \mathbf{B} \rangle| \sim 1)$. However, even when strong assumptions about anisotropy are made (as in ~\citep{Stawarz2009ApJ}), the results have been quite comparable with the isotropic case.

For MMS spacecraft data, the field's components are given in the 
cartesian $GSE$
reference frame. Note that,
since the wind speed in the spacecraft frame is several times larger
than the typical velocity fluctuations and it is nearly
aligned with the $R$ radial direction, time and space lags (scales) ($\tau$ and $r$ respectively) are related approximately through the Taylor hypothesis, $r \approx -\langle V_{x} \rangle_{t} \tau$ (note the sign). From the time series $\mathbf{Z^{\pm}}(t)$, we compute the time
increments $\Delta \mathbf{Z^{\pm}}(\tau ; t) = \mathbf{Z^{\pm}}(t+\tau) - \mathbf{Z^{\pm}}(t)$ and obtain the
mixed third-order structure function $Y^{\pm}(-\langle V_{x} \rangle_{t} \tau) = \langle |\Delta \mathbf{Z^{\pm}}(\tau;t)|^2 \Delta Z_{x}^{\mp}(\tau;t) \rangle_{t}$.
Note that to avoid confusion, 
here moving averages are designated by the notation $\langle \cdots \rangle_{t}$. 

\begin{figure}
	\begin{center}
		\includegraphics[scale=1.0]{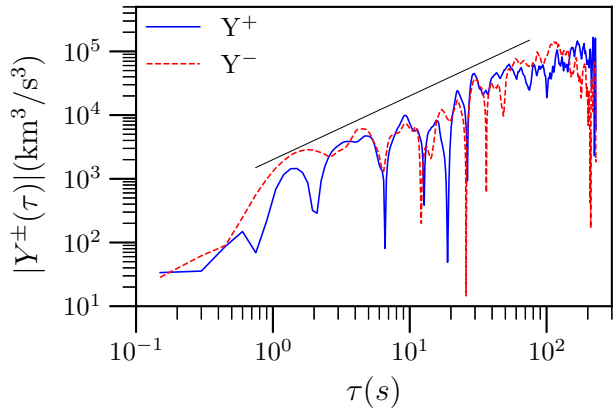}
		\caption{Absolute values of the mixed third-order structure functions, derived from measurements by MMS1. A linear scaling is shown for reference.}
		\label{fig:ypm}
	\end{center}
\end{figure}

In Figure \ref{fig:ypm}, 
we have plotted the absolute values of the mixed third-order structure functions for MMS1. A linear scaling is indeed observed and interpreted here 
according to 
$|Y^{\pm}(\tau)| = (4/3)|\epsilon^{\pm}|\langle V_{x} \rangle_{t}\tau$.
We call this $\epsilon_{2}$. The approximations represented by inserting the 
absolute value will be discussed below. The precise derivation of the signed third-order law for MHD is due to~\citet{Politano1998GRL}. 
By fitting a straight line in the inertial range we obtain for MMS1,
\begin{eqnarray}
|Y^{+}(\tau)| &\simeq& 1014 \tau,\\
\frac{4}{3} \epsilon_{2}^{+} \langle v_{x} \rangle_{t}\tau &\simeq& 1014 \tau,\nonumber\\
\epsilon_{2}^{+} \simeq 5.6 &\times& 10^{6} \mathrm{~J~kg^{-1}~s^{-1}}.
\end{eqnarray}
We follow the same procedure for all four spacecraft and obtain the values listed in Table~\ref{tab:eps2}.

\begin{table}
	\caption{Inertial range cascade rate estimates}
	\label{tab:eps2}
	\begin{center}
	\begin{tabular}{c c c}
		\hline \hline
		S/C 
		& $ \epsilon^{+}_{2} $ & $ \epsilon^{-}_{2} $  
		\\
		& \colhead{($\rm J~kg^{-1}~s^{-1}$)} &  \colhead{($\rm J~kg^{-1}~s^{-1}$)}    
		\\
		\hline
		MMS1 & $5.6\times10^6$ & $3.9\times10^6$  \\
		MMS2 & $7.7\times10^6$ & $5.1\times10^6$  \\
		MMS3 & $7.4\times10^6$ & $5.0\times10^6$  \\
		MMS4 & $7.6\times10^6$ & $5.1\times10^6$  \\
		Average & $(7.1\pm0.9)\times10^6$ & $(4.8\pm0.5)\times10^6$ \\
		\hline
	\end{tabular}
	\end{center}
	\tablecomments{These are based on PP98 MHD adaptation of the Yaglom 
		law using different spacecraft measurements. The last row is the average of all four spacecraft measurements listed in the first four rows. The uncertainty is the standard deviation of the four measurements.}	
\end{table}

The reader should note that there is a reasonable level of agreement among the several independent estimates given in Table~\ref{tab:eps2}. Variability is likely due to poor statistics (use of four samples), anisotropy of the turbulence (equation~(\ref{eq:3ord}) assumes isotropy), and the possibility of energy transfer into a weak compressible cascade (see \cite{Yang2017PoF}).

\section{Kinetic scale: $\epsilon_{3}$} \label{sec:eps3}
At kinetic scales, neither MHD nor 
the incompressive approximation remain valid. 
While total energy conservation remains valid, 
additional dynamical effects 
influence the way that energy is transferred across scales, and converted between forms. 
Influences such as Hall effect, pressure anisotropy and other terms in a generalized Ohm's law need to be considered.
As far as we are aware, so-called ``exact laws'' of the Yaglom type have not been developed for the 
collisionless electron-proton Vlasov plasma. Furthermore, 
even partial descriptions, such as 
compressible Hall MHD,
lead to generalizations of the third order law that 
are quite complex \citep{Andres2018PRE} and include contributions that may be difficult
to evaluate even with the most refined 
observations available. 
Consequently, we adopt a different strategy 
in which we do not attempt a full evaluation of 
a generalized Yaglom law, but only one 
contribution, according to the 
following reasoning, which we present in some detail. 
 
A broad perspective
on the kinetic scale cascade
is that it fragments into multiple channels,
as described, e.g., in~\cite{Yang2017PoP,Yang2017PRE}. 
Using filtering techniques, 
one finds that transfer due to advective nonlinearity of the proton fluid 
persists, at an attenuated level, in the kinetic range, while there are 
also additional channels for energy conversion and transfer. How these
channels of transfer fit into the compact 
picture of a Kolmogorov-Yaglom law requires taking a step back and examining the context in which such laws are derived.

For cases typically considered, ranging from
incompressible hydrodynamics \citep{Kolmogrov1941c} 
to compressible Hall MHD \citep{Andres2018PRE},
one begins with a von Kaman-Howarth equation 
\citep{Karman1938PRSL} written in terms of 
increments with spatial lag $\ell$.
For hydrodynamic turbulence, the latter are longitudinal velocity increments
$\delta u_\ell = \hat \ell \cdot \left [u_\ell({\bf x} + {\bf \ell}) - u_\ell({\bf x})\right ]$ 
while
for incompressible MHD these are longitudinal Elsass\"er increments
$\delta Z_\ell^\pm = \delta u_\ell \pm \delta v_{A\ell}$
where ${\bf v_A}$ is the fluctuation magnetic field in Alfv\'en speed units. 
Following standard manipulations, one arrives
at an equation for the increment energy $E_\ell$
of the form,
\begin{equation}
\frac{dE_\ell}{dt} = 
\nabla_\ell \cdot {\bf Y}
+ \nabla_\ell \cdot {\bf H}
= S + D
\label{eq:3rd} 
\end{equation}
where $\bf Y$ is the vector flux of energy in the inertial range,
$\bf H $ represents a possible additional vector energy flux that 
acts at smaller scales $\ell$, dissipation acting at small $\ell$ 
outside the inertial is represented by $D$, and
other 
sources and sinks outside the inertial range are represented by $S$.
We have taken the liberty of writing
Eq (\ref{eq:3rd}) in a fairly general form
(cf. \citet{Andres2018PRE} and \citet{Hellinger2018ApJL}).

In Eq. (\ref{eq:3rd}), for stationary 
incompressible hydrodynamic turbulence, 
the time derivative vanishes, 
${\bf H} = 0$, $D$ is negligible in the inertial range, 
an $D \to -4 \epsilon$, the total steady dissipation rate, and the 
von Karman equation reduces to the Kolmogorov-Yaglom law
$\nabla_\ell \cdot {\bf Y} = -4\epsilon$.
For compressible hydrodynamics, the internal energy is a new ingredient in 
Eq. (\ref{eq:3rd}), 
and a dilatation ($\nabla \cdot u$)
related 
source $S$ appears in the relation \citep{Banerjee2013PRE}.
For incompressible MHD, the Hall and other kinetic contributions are absent (${\bf H} = 0$), 
and 
 Eq. (\ref{eq:3rd})
becomes the Politano-Pouquet relation
\citep{Politano1998GRL,Politano1998PRE}, wherein 
for steady 
high Reynolds numbers and in the inertial range,
the exact relation 
for mixed third order 
Elsasser correlations emerges as 
$\nabla_\ell  \cdot {\bf Y} = \nabla_\ell \cdot \langle \delta Z_\ell^\mp 
|\delta Z_\ell^\pm \rangle = -4 \epsilon^\pm$.

Correspondingly, for incompressible
Hall MHD \citep{Galtier2008JGR}, ${\bf H}$ is due to 
the Hall effect and of order $d_i/L$, inertial length $d_i$ and 
energy containing scale $L$.
For this case, the ``Yaglom flux'' ${\bf Y}$ 
remains the dominant contribution
when $\ell \ll L$ but still in the inertial range,
and vector flux $\bf Y$ remains as in the MHD case, 
while at scales $\ell \sim d_i$ and smaller the 
Hall contribution $\bf H$ becomes more
important and in this range of scales, the 
exact law becomes 
$\nabla_\ell \cdot ({\bf Y} + {\bf H}) = -4\epsilon$.
The standard MHD vector flux remains, but contributes 
at a diminished level.

This is an important feature 
of the Yaglom-like third-order models that may not always be 
fully appreciated.
When moving 
outside of the range of strict applicability of 
the simplest form $\nabla_\ell \cdot {\bf Y} = -4\epsilon$,
i.e., outside of the range of the exact law,
the more general von Karman relation Eq. (\ref{eq:3rd}), still holds, as 
additional terms begin to 
make significant contributions.
This approach was adopted recently by 
\citet{Hellinger2018ApJL}
who examined energy transfer in a hybrid Vlasov 
(particle-in-cell ions; fluid electrons) 
employing an incompressible 
Hall MHD formalism. 
As suggested above, the 
relatively more complex 
description of energy transfer in 
compressible Hall MHD \citep{Andres2018PRE}
involves relatively complex 
new source terms ($S$) that may be difficult to evaluate.

Here 
we will examine transfer at subproton scales in the magnetosheath using MMS data.
We will examine only the Politano-Pouquet
energy flux in its general divergence form 
to arrive at a partial estimate of the 
total cascade at those scales. 
In particular, 
we exploit the 
small MMS inter-spacecraft separation to 
carry out a direct evaluation
that has not been previously possible.

For the present interval, average separation between MMS2 and MMS4 is about 7.16 km which is intermediate between $d_i=17.5$ km and $d_e = 0.4$ km. 
Energy cascade at these small 
scales is expected to be well into the kinetic regime and may not be described well by MHD inertial scale phenomenologies.
Nevertheless, we may estimate 
this   
contribution to energy transfer, say $\epsilon_3^\pm$ 
making 
formal use of 
the Politano-Pouquet 
estimate of the incompressive
energy vector flux. Thus,  
in terms of the Elsasser variables
$\mathbf{Z}^{\pm}$,
\begin{eqnarray}
\nabla \cdot \langle \Delta \mathbf{Z}^{\mp} |\Delta \mathbf{Z}^{\pm}|^2 \rangle = 
- 4 \epsilon_3^{\pm}\label{eq:ky}
\end{eqnarray}
where we expect that $\epsilon_3^\pm < \epsilon^\pm.$ 
This equation 
does not assume any form of spectral distribution while equation~(\ref{eq:3ord}) requires isotropy. 
This result is implicit in ~\citet{Politano1998GRL}
as has been pointed out by~\citet{MacBride2008ApJ}.
In exchange for this generality, solution of 
equation~(\ref{eq:ky}) is not algebraic, but requires, in effect, 
Gaussian integration over a closed surface, as has been extensively 
discussed in the literature~\citep{Wan2009PoP,Stawarz2009ApJ,Osman2011PRL}.

MMS data, having wide angle coverage along with measurement from four spacecraft, is suitable
for adopting a multi-spacecraft estimate of 
the cascade rate using Gauss's law and 
the more general form 
of the  Kolmogorov-Yaglom law equation~(\ref{eq:ky}),
without assuming isotropy etc. 
By adopting this approach, we also have no need to 
employ the Taylor hypothesis, as we use only two-spacecraft correlation estimates.
Note that the spacecraft separations 
lie in the sub-proton scale kinetic range. 
\begin{figure}
	\begin{center}
		\includegraphics[scale=0.6]{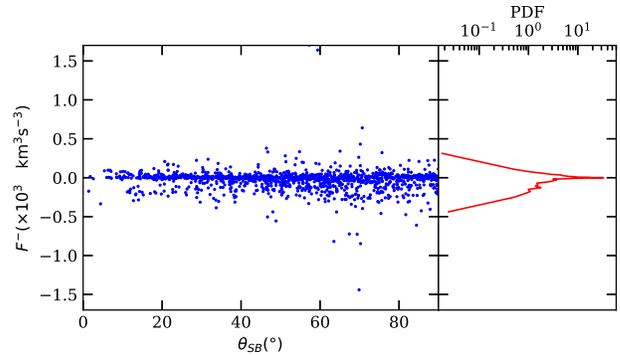}
		\caption{All estimation of flux density with the corresponding PDF for MMS2 and MMS4.}
		\label{fig:Fm}
	\end{center}
\end{figure}
Proceeding accordingly, 
a field angle $\theta_{\mathrm{SB}}$ is defined as the acute angle between the time-lagged spacecraft separation vector and the field direction. Good coverage in $\theta_{\mathrm{SB}}$ is needed to accurately evaluate equation~(\ref{eq:ky}). 
Using 
Gauss's law and integrating over a sphere, we find
\begin{eqnarray}
\int_{0}^{\pi /2} \langle F^{\pm} \rangle \sin \theta_{\mathrm{SB}} d \theta_{\mathrm{SB}} = - \frac{4}{3} \epsilon^{\pm} |\mathbf{r}|\label{eq:calc_eps3},
\end{eqnarray}
where $F^{\pm}= (\mathbf{\hat{r}} \cdot \Delta \mathbf{Z}^{\mp}) |\Delta \mathbf{Z}^{\pm}|^2$ is the flux density. 
We call this estimate of the cross-scale 
energy transfer rate $\epsilon_{3}$.
MMS has four spacecraft, so a total of six pairs are possible to evaluate equation~(\ref{eq:calc_eps3}). Each pair has slightly different separation, so we calculate the left-hand-side of equation~(\ref{eq:calc_eps3}) and $\mathbf{r}$ separately for each pair. An example using the two spacecraft MMS2 and MMS4, is shown in Fig.~\ref{fig:Fm}. The data is then binned and averaged. $d \theta_{\mathrm{SB}}$ in equation~(\ref{eq:calc_eps3}) corresponds to the bin widths while $\langle F^{\pm} \rangle$ corresponds to arithmetic mean of $F^{\pm}$ within a bin.
We report values obtained from equation~(\ref{eq:calc_eps3}) for all combinations of spacecraft pairs in Table~\ref{tab:eps3}.

\begin{table}
	\caption{$\epsilon_{3}$ for different spacecraft pairs}
	\label{tab:eps3}
	\begin{center}
	\begin{tabular}{c c c c}
		\hline \hline
		S/C pair
		& $|\mathbf{r}|$
		& $ \epsilon^{+}_{3} $
		& $ \epsilon^{-}_{3} $  
		\\
		 & \colhead{($\rm km$)} &  \colhead{($\rm J~kg^{-1}~s^{-1}$)} & \colhead{($\rm J~kg^{-1}~s^{-1}$)}  
		\\
		\hline
		1-2 &  $8.252 \pm 0.003$ & $ 0.95 \times 10^{6}$ & $ 0.34 \times 10^{6}$ \\
		1-3 &  $8.490 \pm 0.005$ & $ 0.35 \times 10^{6}$ & $ 0.13 \times 10^{6}$ \\
		1-4 &  $9.85 \pm 0.02$ & $ 0.66 \times 10^{6}$ & $ 0.71 \times 10^{6}$ \\
		2-3 &  $6.435 \pm 0.004$ & $ 0.53 \times 10^{6}$ & $ 1.07 \times 10^{6}$ \\
		2-4 &  $7.169 \pm 0.007$ & $ 2.49 \times 10^{6}$ & $ 3.34 \times 10^{6}$ \\
		3-4 &  $8.02 \pm 0.02$ & $ 1.04 \times 10^{6}$ & $ 3.10 \times 10^{6}$ \\
		Average & $--$ & $ (1.0 \pm 0.7) \times 10^{6}$ & $(1\pm1) \times 10^{6}$ \\
		\hline
	\end{tabular}
	\end{center}
	\tablecomments{The first column represents the spacecraft pairs used for study. The last row is the average of all of the previous rows. The uncertainty has been evaluated by calculating the standard deviation.}
\end{table}

\section{Conclusions} \label{sec:conc}

In this study, we have employed the special 
characteristics of the MMS spacecraft and instrumentation 
to provide 
distinct estimates of the cascade rate using three methodologies
that span a wide range of scales.  
Using the MHD extension of the the von K\'arm\'an decay law, 
the decay rate at energy-containing scales is 
estimated in magnetosheath spacecraft observations, for the first time,
as far as we are aware. 
The Politano-Pouquet 
third-order law provides the basis for an inertial range cascade rate estimate.
Finally, 
a multi-spacecraft technique has been used 
at the kinetic scale, also we believe for the first time, to estimate 
the (partial) energy transfer transfer rate 
via incompressive channel using the Kolmogorov-Yaglom law.

In Table~\ref{tab:eps123}, we list the average values of $\epsilon^{\pm}$, estimated
using different methods in three ranges of scale. 
\begin{table}
	\caption{Estimation of cascade rate at different scales}
	\label{tab:eps123}
	\centering
	\begin{tabular}{c c c c}
		\hline \hline
		 & $ \epsilon_{1} $ 
		 & $ \epsilon_{2} $
		 & $ \epsilon_{3} $  
		\\
		  & \colhead{($\rm J~kg^{-1}~s^{-1}$)} &  \colhead{($\rm J~kg^{-1}~s^{-1}$)} & \colhead{($\rm J~kg^{-1}~s^{-1}$)}  
		\\
		\hline
		$\epsilon^{+}$ & $ (5.8\pm0.5) \times 10^{6}$ & $ (7.1\pm0.9) \times 10^{6}$ & $ (1.0\pm0.7) \times 10^{6}$ \\
		$\epsilon^{-}$ & $ (5.6\pm0.2) \times 10^{6}$ & $ (4.8\pm0.5) \times 10^{6}$ & $ (1\pm1) \times 10^{6}$ \\
		\hline
	\end{tabular}
\end{table}
It can be seen from Table~\ref{tab:eps123} that the decay rate, obtained from von K\'arm\'an decay phenomenology and the inertial range cascade evaluated from the third-order law, are in agreement with each other within uncertainties. As discussed in the main text, there are several choices for the similarity lengths and the proportionality constants in the von K\`arm\'an decay law. The results presented here indicate that the conventions adopted here are probably appropriate.

The cascade rate evaluated at the kinetic range, using multi-spacecraft method is 
lower than the inertial range and the von K\'arm\'an decay rate. This is 
expected, since in the kinetic 
range additional channels, not present in single-fluid MHD, 
open up for energy conversion and transfer, as described
in recent theoretical works~\citep{Howes2008PoP,Howes2008JGR,Del_Sarto2016,Yang2017PoP,Yang2017PRE}. These additional
pathways may be associated with wave-particle interactions, 
kinetic activity related to reconnection, compressive and incompressive cascades, distinct cascades 
for different species, and so on.  This is a much more complex scenario 
than a single incompressible Kolmogorov cascade, which is often the standard 
viewpoint at MHD scales. 
The fact that 
$\epsilon_{1} \sim \epsilon_{2} > \epsilon_3$ 
demonstrates empirically that
the standard Kolmogorov cascade may be operative in the kinetic scales, 
as an ingredient of a more complex cascade, and therefore at a diminished intensity. 
A more rigorous approach would be to derive the appropriate third-order law relevant to the kinetic scale plasma turbulent as described in the literature~\citep{Schekochihin2009ApJS,Boldyrev2013ApJ,Kunz2018JPP,Eyink2018}. 
A more complete statistical study of a large sample of data is required to confirm such conclusions. We are in the process of performing similar study with 
a wider variety of datasets.

Another interesting observation from Table~\ref{tab:eps3}
is that although the spacecraft separations $|\mathbf{r}|$ are almost equal, the cascade rates are quite widely distributed. As discussed before, Eq.~(\ref{eq:ky}) does not assume isotropy. Therefore the spread in cascade rate for different spacecraft pairs may be a result of small scale inhomogeneity and anisotropy becoming progressively stronger as smaller scales are probed, as previously investigated in MHD systems~\citep{Shebalin1983JPP,Oughton1994JFM,Milano2001PoP}. 
This is beyond the scope of the present paper, but we 
wish to address this issue in the future. 
Also, for each case, $\epsilon^{+}$ and $\epsilon^{-}$ lie within each other's uncertainty limits, which is expected, because cross helicity is very low for the selected interval.
 
Finally, we note the comparison of 
estimated cascade rates
in the magnetosheath and in the solar wind. 
For convenience, we designate 
the incompressible cascade rate in the 
magnetosheath as $\epsilon_{\mathrm{MSH}}$,
and the corresponding rate in the pristine solar wind as $\epsilon_{\mathrm{SW}}$. 
We only consider nearly incompressible, nearly isotropic, low cross helicity plasma, as
varying these conditions might change the situation. Previously,~\cite{Hadid2018PRL} (HadidEA hereafter)  
also calculated the cascade rate in 
Alfv\'enic incompressible magnetosheath turbulence, finding, for 
estimates in units of energy/volume, 
\begin{eqnarray}
\frac{\epsilon_{\mathrm{MSH}}}{\epsilon_{\mathrm{SW}}}\bigg|_{\mathrm{HadidEA}} \simeq \frac{10^{-13}\mathrm{~J~m^{-3}~s^{-1}}}{10^{-16}\mathrm{~J~m^{-3}~s^{-1}}} = 10^{3}.
\end{eqnarray}
From our analysis, in units of energy/mass, we find 
\begin{eqnarray}
\frac{\epsilon_{\mathrm{MSH}}}{\epsilon_{\mathrm{SW}}}\bigg|_{\mathrm{we}} \simeq \frac{5 \times 10^{6}\mathrm{~J~kg^{-1}~s^{-1}}}{50 \times 10^{2}\mathrm{~J~kg^{-1}~s^{-1}}} \simeq 10^{3},
\end{eqnarray}
where we use the~\cite{Osman2011PRL} value for $\epsilon_{\mathrm{SW}}
 \simeq 50 \times 10^{2}\mathrm{~J~kg^{-1}~s^{-1}}$.

On the basis of the above results, 
and in accord with ~\cite{Hadid2018PRL}, 
we conclude that the Earth's magnetosheath 
cascade rate is much higher than that of the solar wind, 
even if the observed turbulence is quasi-incompressive.
Recall that the  
density fluctuations are rather small (Table~\ref{tab:overview}).
This high cascade rate is presumably due to 
strong driving of the magnetosheath by the 
solar wind through compressions and streaming through the bow shock,
amplifying the preexisting turbulence activity.  
Due to the nature of this driving, the magnetosheath 
is nominally more turbulent and hotter than the nearby 
solar wind, while the high value of plasma beta 
allows the turbulent Mach number to remain relatively 
low (see Table \ref{tab:overview}).
From a theoretical point of view, 
this places magnetosheath turbulence in a somewhat different
category than pristine solar wind turbulence is in.
Understanding their relationship provides an interesting 
further challenge for plasma turbulence theory. 
In this regard, it would be interesting to compare the findings of this paper with other cases with high compressibility, high cross helicity etc. We plan to perform these studies in 
the future.

\section*{Acknowledgments}
The authors are grateful to Hugo Breuillard for providing valuable comments and references. 
This research partially supported by the MMS Mission
through NASA grant NNX14AC39G at the University of
Delaware, by NASA LWS grant NNX17AB79G, and by the 
Parker Solar Probe Plus project through Princeton/ISOIS 
subcontract SUB0000165,
and in part
by NSF-SHINE AGS-1460130. A.C. is supported by the
NASA grants. W.H.M. is a member of the MMS Theory and
Modeling team. We are grateful to the MMS instrument teams,
especially SDC, FPI, and FIELDS, for cooperation and
collaboration in preparing the data. The data used in this
analysis are Level 2 FIELDS and FPI data products, in
cooperation with the instrument teams and in accordance their
guidelines. All MMS data are available at \url{ https://lasp.colorado.edu/mms/sdc/}.


\end{document}